\documentclass[%
 reprint,
showkeys,
 amsmath,amssymb,
 aps,
]{revtex4-1}

\usepackage{float}
\usepackage{graphicx}
\usepackage{dcolumn}
\usepackage{bm}
\usepackage[T1]{fontenc}
\usepackage{textcomp}
\usepackage[outercaption]{sidecap}

\begin{document}

\preprint{APS/123-QED}

\title{$^{57}$Fe M\"ossbauer study of Lu$_2$Fe$_3$Si$_5$ iron silicide superconductor}

\author{Xiaoming Ma$^{1,2}$, Sheng Ran$^{2}$, Hua Pang$^{1}$, Fashen Li$^{1}$ Paul C. Canfield$^{2}$ and Sergey L. Bud'ko$^{2}$}
\email{Corresponding author: budko@ameslab.gov}

\affiliation{ $^1$Institute of Applied Magnetics, Key Laboratory for Magnetism and Magnetic Materials of the Ministry of Education, Lanzhou University, Lanzhou, Gansu Province, 730000, China.\\$^2$Ames Laboratory, US DOE, and Department of Physics and Astronomy, Iowa State University, Ames, Iowa 50011, USA}

\date{\today}

\begin{abstract}
With the advent of Fe-As based superconductivity it has become important to study how superconductivity manifests itself in details of $^{57}$Fe M\"ossbauer spectroscopy of conventional, Fe - bearing superconductors. To this end, the iron-based superconductor Lu$_2$Fe$_3$Si$_5$ has been studied by $^{57}$Fe M\"ossbauer spectroscopy over the temperature range from 4.4 K to room temperature with particular attention to the region close to the superconducting transition temperature (T$_c$ = 6.1 K). Consistent with the two crystallographic sites for Fe in this structure, the observed spectra appear to have a pattern consisting of two doublets over the whole temperature range. The value of Debye temperature was estimated from temperature dependence of the isomer shift and the total spectral area and compared with the specific heat capacity data. Neither abnormal behavior of the hyperfine parameters at or near T$_c$, nor  phonon softening were observed.
\begin{description}
\item[PACS numbers]74.70.Dd, 76.80.+y
\end{description}
\end{abstract}

\pacs{Valid PACS appear here}
\maketitle


\section{\label{sec:level1}Introduction}
Since the discovery of M\"ossbauer effect spectroscopy more than half a century ago \cite{ADD1_ZP1958}, superconductivity has been one of the states that has been investigated by this technique \cite{ADD3_NC1961}. Although M\"ossbauer spectroscopy is widely accepted as one of the most sensitive techniques in terms of energy resolution, it has not contributed significant insight to studies of conventional superconductors. After the discovery of cuprate high temperature superconductors (HTSC), M\"ossbauer spectroscopy was widely used for studies of these materials and reports of observation of some anomalies in the spectral parameters in the vicinity of the superconducting critical temperature (T$_c$)  \cite{1PRL1991,2PRB1992,3PRB1988,4JPCM1993,5JPCM1994} were published. Due to the absence of commonly available M\"ossbauer nuclides in the cuprates, most studies were accomplished either by partial substitution of copper atoms by $^{57}$Fe and/or $^{119}$Sn, or by using resonant isotopes of the rare earth metals, like $^{151}$Eu, which increases the degree of difficulty of the measurements and reduces the clarity of the results \cite{1PRL1991,3PRB1988,6SSC1987,7PC1999}. Recently, the discovery of  iron-based superconductors, that naturally contain the common M\"ossbauer nuclide, $^{57}$Fe, has triggered intense M\"ossbauer studies of these superconductors \cite{8PC2009,9PRB2011,10EPL2014,11PRB2011,12HI2013,13HI2012}. Superconductivity in iron-pnictides is usually achieved by doping a magnetic parent compound with electrons or holes, or by application of chemical or physical pressure, and thereby, suppressing the magnetic order, suggesting that superconductivity and magnetism are closely related in this system. Although there are some studies on iron-based superconductors, which state that M\"ossbauer spectral parameters show anomalies near T$_c$, in these materials it is  hard to attribute the variation of the hyperfine parameters observed by M\"ossbauer spectroscopy to purely magnetic or purely superconducting origins \cite{10EPL2014,12HI2013,13HI2012,14PRB2013}.

In order to study possible variations of hyperfine parameters caused by the transition from the normal to the superconducting state in a conventional superconductor, we revisited lutetium-iron-silicide, Lu$_2$Fe$_3$Si$_5$. Lu$_2$Fe$_3$Si$_5$ is a stoichiometric, Fe-containing, superconductor with relatively high T$_c \approx$ 6 K \cite{15PL1980}. In a previous  M\"ossbauer study \cite{16JMMM1981}, the non-magnetic nature of Fe was already confirmed.  Hence,  Lu$_2$Fe$_3$Si$_5$ can be considered as an ideal compound to investigate the variation of hyperfine parameters caused only by the superconducting transition without any complications associated by the absence of M\"ossbauer nucleus and/or the presence of magnetism. To the best of our knowledge no detailed, temperature dependent, $^{57}$Fe M\"ossbauer spectroscopy measurements were performed on this material so far, and our goal is to shed some light on the applicability of M\"ossbauer spectroscopy for studies of conventional, albeit multigap superconductors \cite{ADD2_PC2012}.

\section{Experimental details}
Polycrystalline samples of Lu$_2$Fe$_3$Si$_5$ were prepared by arc melting constituent elements with the nominal composition of Lu$_{2}$Fe$_{3.32}$Si$_{5.26}$ (corresponding to Lu$_2$Fe$_3$Si$_5$ + Fe$_{0.32}$Si$_{0.26}$) in Zr-gettered Ar atmosphere. Extra iron was added to suppress the formation of a  Lu$_2$FeSi$_4$ second phase and extra silicon was added to compensate apparent loss during the arc melting. To ensure the homogeneity of the sample, the arc melting was repeated iteratively after flipping of the melted and resolidified ingot, for more than ten times. The weight loss was about 0.26\%. The arc-melted ingot was then sealed in an amorphous silica tube, under a partial pressure of argon, and annealed at 1050 \textcelsius \ for 12 days.

Powder X-ray diffraction (XRD) was performed using a Rigaku Miniflex diffractometer with Cu K$\alpha$ radiation at room temperature (RT). The powder X-ray spectra of the samples were refined by Rietveld analysis using the EXPGUI software \cite{17EXPGUI}. Magnetic measurements were performed using a Quantum Design Magnetic Property Measurement System SQUID magnetometer, specific heat capacity was measured in a Quantum Design Physical Property Measurement System.

M\"ossbauer spectroscopy measurements were performed using a SEE Co. conventional constant acceleration type spectrometer in transmission geometry with  an $^{57}$Co(Rh) source, which had an initial intensity 50 mCi, kept at RT. The absorber was prepared in a powder form (10 mg of natural Fe/cm$^2$) by grinding of approximately 175 mg piece of the arc-melted and annealed button. The absorber holder comprised two nested white Delrin cups. The powder was placed uniformly on the bottom of the larger cup and was held in place by a smaller cup. The absorber holder was locked in a thermal contact with a copper block with a temperature sensor and a heater, and aligned with the $\gamma$ - source and detector.  The absorber was cooled to a desired temperature using Janis model SHI-850-5 closed cycle refrigerator (with vibrations damping) that has long-term temperature stability better than 0.1 K at low temperature. The driver velocity was calibrated by $\alpha$-Fe foil and all isomer shifts (IS) are quoted relative to the $\alpha$-Fe foil at RT. At first,  spectra with maximum velocity 6 mm/s and 3 mm/s were both measured at RT to check that no iron-containing impurity can be seen in the M\"ossbauer spectra. Then, three rounds of measurements, progressively focusing in on temperatures near T$_c = 6.1$ K, were carried out: collecting 24 h with maximum velocity 2 mm/s from 4.3 K to 293.8 K (S1); collecting 48 h with maximum velocity 3 mm/s from 4.4 K to 10 K (S2); collecting 48 h with maximum velocity 3 mm/s from 4.7 K to 6.4 K (S3). All the M\"ossbauer spectra were fitted by the commercial software package MossWinn \cite{18MossWinn}, in which the standard error of parameters can be estimated either by calculating and inverting the curvature matrix of the $\chi^2$ with respect to the fit parameters, or by the Monte Carlo method. The standard error of IS, quadroupole splitting (QS) and line width ($\Gamma$) in this work were obtained from the curvature matrix, while the error the area under the spectra here were obtained by Monte Carlo method by iterating 100 times.

\begin{figure}[htp]
\centering
\includegraphics[width=0.5\textwidth]{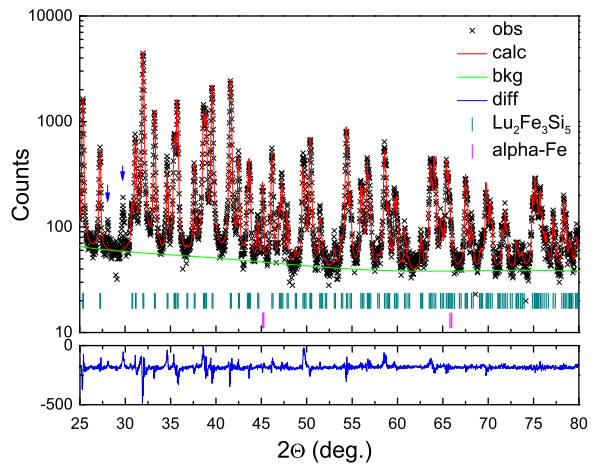}
\centering
\caption{\label{fig:epsart} (Color online) Rietveld refinement of the powder XRD spectrum of Lu$_2$Fe$_3$Si$_5$. Measured (black cross), calculated intensities (red line) and background (green line) and difference curve (blue line) are shown. Vertical bars at the bottom indicate the positions of the Bragg reflections. The peaks of unknown phase are marked by the blue arrows.}
\end{figure}

\begin{figure}[htp]
\includegraphics[width=0.5\textwidth]{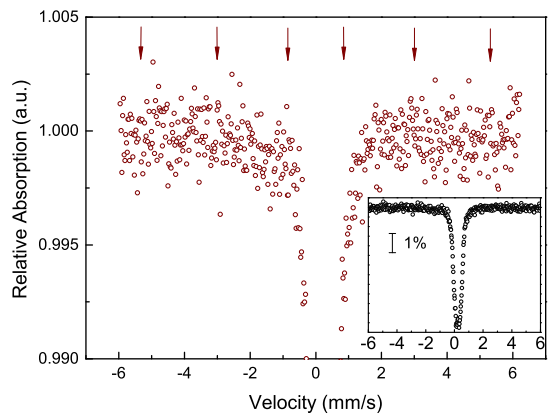}
\caption{\label{fig:epsart}(Color online) Expanded view of the background part of the RT Lu$_2$Fe$_3$Si$_5$ M\"ossbauer spectrum in large velocity scale. Red arrows show the expected peaks positions for $\alpha$-Fe.The inset is the full view of the spectrum.}
\end{figure}

\section{Results and discussion}

\begin{figure}[htp]
\centering
\includegraphics[width=0.48\textwidth]{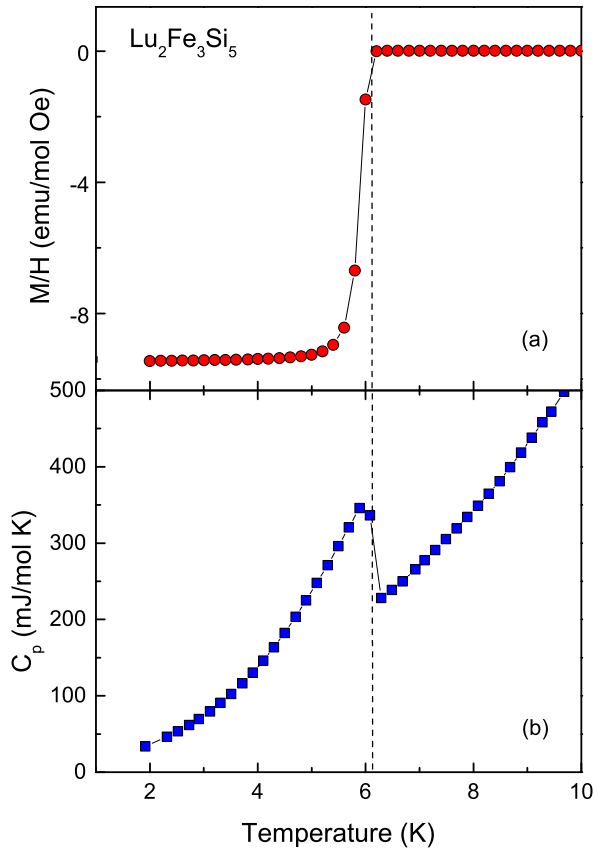}
\centering
\caption{\label{fig:epsart} (Color online) (a) Temperature dependent of zero field cooled magnetic susceptibility of Lu$_2$Fe$_3$Si$_5$ (H = 10 Oe); (b) the temperature dependent low temperature specific heat capacity. Vertical dashed line marks T$_c = 6.1$ K.}
\end{figure}

\begin{figure}[htp]
\centering
\includegraphics[width=0.5\textwidth]{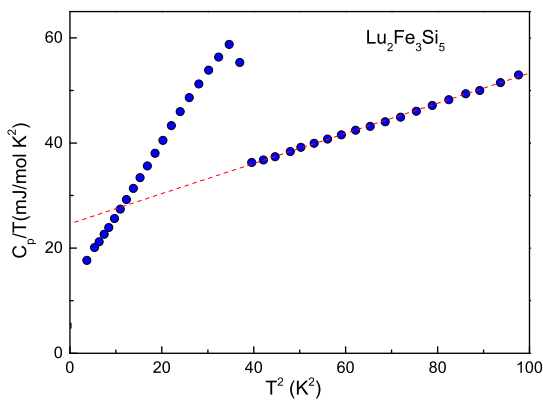}
\centering
\caption{\label{fig:epsart} (Color online) The specific heat capacity divided by temperature, C$_p$/T, as a function of T$^2$ for Lu$_2$Fe$_3$Si$_5$. The dashed line represents the linear fit to the data in the normal state as described in the text.}
\end{figure}

\subsection{Structure and superconductivity}
The XRD pattern of the Lu$_2$Fe$_3$Si$_5$ polycrystalline sample is presented in Fig. 1. The majority of the peaks match to the tetragonal structure with the $P4/mnc$ space group. The Rietveld refinement results in an estimate  of $\sim 1.8(1)$ wt.\% $\alpha$-Fe impurity and a trace amount of unknown phase. A 1.8 wt.\% of $\alpha$-Fe  corresponds to 6.7 \% of Fe atoms in the $\alpha$-Fe form in the ground sample, which, for un-enriched Fe, is below the resolution of $^{57}$Fe M\"ossbauer spectroscopy to detect $\alpha$-Fe. As can be seen in Fig. 2, there are no peaks associated with the $\alpha$-Fe or with the other impurity in the M\"ossbauer spectrum, which means the unknown phase is either iron-free or, if contains iron, is below the resolution limit. Consequently, the M\"ossbauer spectra can be analyzed as a single - phase (Lu$_2$Fe$_3$Si$_5$) spectra.

The superconductivity of the sample is confirmed by the dc susceptibility measurement in a magnetic field of 10 Oe. As shown in Fig. 3 (a), the susceptibility data shows diamagnetic signal below $\sim 6.2$ K. The transition is sharp with a width of less than 0.4 K. Fig. 3 (b) shows the temperature dependent specific heat capacity (C$_p$). A sudden jump caused by superconducting transition can be observed below 6.3 K on cooling, the transition temperature is close to that obtained from the susceptibility data. Based on these two thermodynamic measurements we take T$_c$ = 6.1$\pm$0.1 K.

C$_p$/T is also plotted as a function of T$^2$ in Fig. 4. A linear fit above the superconducting transition yields the values of $\gamma_n$ ($\gamma_n$T is the electronic contribution to specific heat capacity) and $\beta_n$ ($\beta_n$$T^3$ represents the phonon contribution) of $\gamma_n$ = 24.6(2) mJ/mol K$^2$, $\beta_n$ = 0.287(2) mJ/mol K$^4$, which are very close to the previously reported values \cite{19PRL2008}. From the $\beta$ value, we can estimate the value of the Debye temperature ($\Theta$$_D$) using the relation: $\Theta$$_D$ = (12$\pi$$^4$$N$$r$k$_B$/5$\beta$)$^{1/3}$, where $N$ is Avogadro's number, $r$ is the number of atoms per formula unit, and k$_B$ is the Boltzmann's constant. We further obtained $\Theta$$_D$ is 408 K. The value of the normalized specific-heat jump at T$_c$, $\Delta$C/$\gamma_n$T$_c$, is $\approx 1.06$, a value that is smaller than the BCS value of 1.43, but consistent with the previously reported value 1.05 \cite{19PRL2008}. From the above characterizations, we can conclude that our sample is a bulk superconductor and can be considered as single phase for M\"ossbauer measurements and analysis.

\subsection{M\"ossbauer results and discussion}
\subsubsection{Symmetry of Fe sites and choice of the model}
In the Lu$_2$Fe$_3$Si$_5$ crystal structure, there are two, nonequivalent, Fe positions, Fe$_{\rm{I}}$ and Fe$_{\rm{II}}$ of 1 : 2 occupation. Fe$_{\rm{I}}$ atoms are located at the 4d sites, which form 1D chains along the $c$ axis.  Fe$_{\rm{II}}$ atoms are located at 8h sites, which form squares with planes perpendicular to the $c$ axis. In each Fe position, the Fe atom is located in a polyhedron formed by Si atoms. Fe$_{\rm{I}}$ has four Si atoms at a distance of 2.31 \AA\ which form an irregular tetrahedron and two Si atoms at a distance of 2.54 \AA\ with the nearest Fe-Fe distance is 2.67 \AA. The Fe$_{\rm{II}}$ has four Si atoms at a distance of 2.34 \AA\, which are in the same face and form a quadrangle. On each side of the face, there are two Si atoms with the 2.35 \AA\ Fe-Si distances. The nearest Fe-Fe distance for Fe$_{\rm{II}}$ is 2.71 \AA.

The anisotropic environments of the Fe atoms ensure nonzero electric field gradient (EFG) tensor at both sites. Hence, in analyzing the data, two doublets are expected. All the $^{57}$Fe M\"ossbauer spectra of Lu$_2$Fe$_3$Si$_5$ over the whole observed temperature range share similar spectral shapes with a clear quadrupole splitting as shown in Fig. 5 (a). A very small asymmetry and small shoulders were observed, which suggest that at least two subspectra are needed to resolve the spectra. However, due to the small splitting and consequently poor resolution of the spectra, the data can be analyzed with more than one set of parameters. In the previous M\"ossbauer studies of R$_2$Fe$_3$Si$_5$ (R = rare earth), both (i) one doublet, and (ii) two doublets with fixed area ratio were employed to fit the spectra \cite{16JMMM1981,20JAP1981}. In our approach to the fitting, two doublets without any restriction are used. In model 1, the two subspectra have close values of QS but obviously different values of IS, which is similar to the reported Sc$_2$Fe$_3$Si$_5$ fit result \cite{21PL1980}. In model 2, the two subspectra have similar values of IS, but distinct QS values. Two sets of parameters yielding fits of acceptable quality can be obtained, the corresponding  parameters of the RT spectrum fits are listed in the  Table 1. The RT spectrum fitted using our two models is shown in Fig. 5 (b) and (c). The relative area of two subspectra in  model 2 are closer to the theoretical value 1 : 2 and $\chi^2$ values are closer to 1, so we have chosen the model 2 to fit all the collected spectra. As an example, the fit using  model 2 of the spectrum measured at 4.4 K from the S2 set is shown in Fig. 5 (d) and the corresponding parameters are listed in Table 1.

\begin{figure}[htp]
\includegraphics[width=0.48\textwidth]{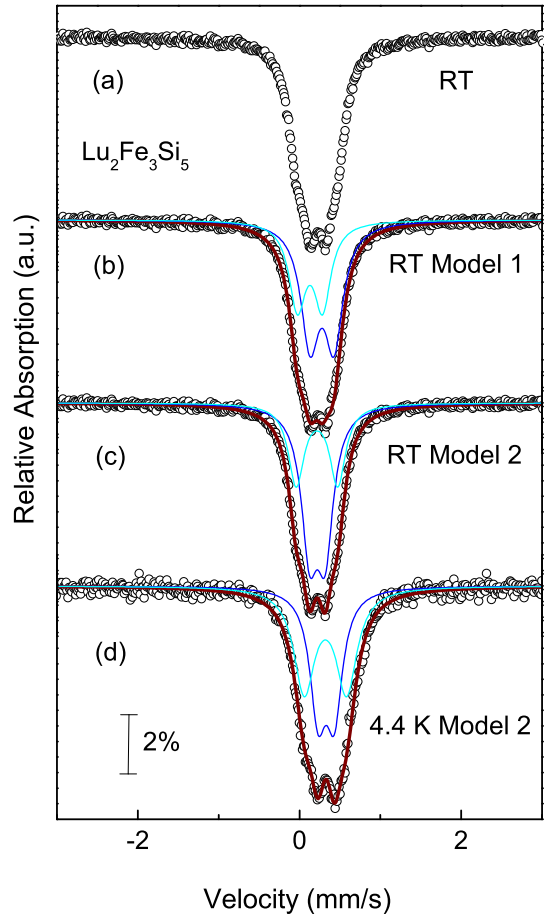}
\caption{\label{fig:wide}(Color online) $^{57}$Fe M\"ossbauer spectra of Lu$_2$Fe$_3$Si$_5$. (a)the spectrum at RT; the results of the fits using model 1 and model 2 are shown in (b) and (c), respectively; (d) is the spectra collected at 4.4 K and fitted using the model 2. The  model 1 and model 2 are described in detail in the text.}
\end{figure}

\begin{table}
\caption{\label{tab:table4}
Hyperfine parameters obtained by fitting using different models as discussed in the text. IS is the isomer shift, QS is the quadrupole splitting and I is the relative area of the two subspectra. }

\begin{ruledtabular}
\begin{tabular}{ccccccc}
T          &model      &site    &IS         &QS         &I            &$\chi^2$  \\
\mbox{K}   &           &        &\mbox{mm/s}&\mbox{mm/s}&\mbox{\%}    &          \\
\hline
RT         &1          &4d      & 0.129(3)  &0.311(1)   &39.4         &1.63      \\
           &           &8h      & 0.278(3)  &0.296(1)    &60.6         &          \\
RT         &2          &4d      & 0.2157(9) &0.513(6)   &36.4         &0.98      \\
           &           &8h      & 0.2206(6) &0.248(8)   &63.6         &          \\
4.4        &2          &4d      & 0.320(2)  &0.523(4)   &52.8         &1.29      \\
           &           &8h      & 0.330(1)  &0.202(3)   &47.2         &          \\
\end{tabular}
\end{ruledtabular}
\end{table}

\begin{figure*}[htp]
\includegraphics[width=0.7\textwidth]{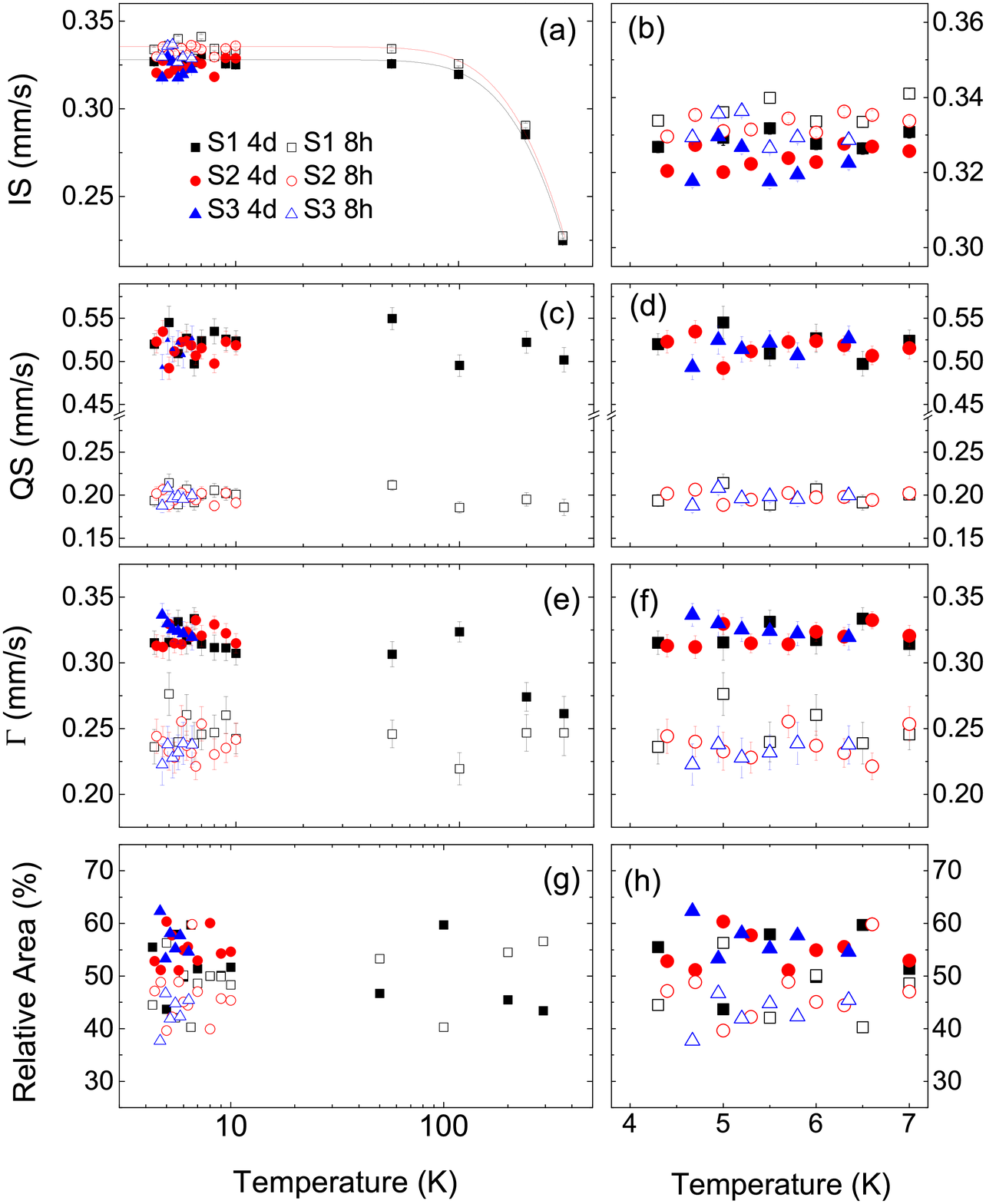}
\caption{\label{fig:epsart}(Color online) Temperature dependence of parameters derived from fitting the of Lu$_2$Fe$_3$Si$_5$ M\"ossbauer spectra with model 2 as described in the text. (a) Isomer shift, (c) quadrupole splitting, (e) line width and (g) relative area of two subspectra. S1, S2, S3 - mark three measurement sets, the parameters are given for two Fe sites, 4d and 8h (see the text for more details).  The lines in panel (a) are fits to the Eq. (1). The figures in the right column are the expansion one of the low temperature region of the corresponding left figure. The temperature stability for each of the measurements in the right column is better than that presented by the size of the symbols.}
\end{figure*}

\subsubsection{Hyperfine parameters}

Fig. 6 summarizes the variation of hyperfine parameters of the spectra with temperature.  The plots in the right column present an expanded view of the low temperature range.

The IS of Lu$_2$Fe$_3$Si$_5$ plotted as a function of temperature is shown in Fig. 6 (a) and (b). The IS values of the 4d and 8h sites at 294 K are 0.225(2) mm/s and 0.227(1) mm/s, respectively, which are slightly larger than the typical values for iron-silicon compounds in which Fe carries no moment \cite{22JMMM1983,23JPCSSC1973}, but are about 0.2 mm/s smaller than that in the iron-pnictide compounds \cite{8PC2009,9PRB2011,10EPL2014}. The temperature dependencies of the IS corresponding to the two sites are very similar, and no anomalies can be observed around T$_c$ (Fig. 6(b)). The IS values obtained from the fits includes contributions from both the chemical shift and the second-order Doppler shift, which is known to increase convexly upon decreasing temperature, due to gradual depopulation of the excited phonon states. However, it should be constant at low temperature, because of the quantum mechanical zero-point motion. The chemical shift should not depend on temperature. The main contribution to this variation is from the second-order Doppler shift, which is usually described by Debye model:
\begin{equation}
IS(T)=IS(0)-\frac{9}{2}\frac{k_BT}{Mc}(\frac{T}{\Theta_D})^3\int_0^{\Theta_D/T}\frac{x^3dx}{e^x-1},
\end{equation}
where c is the velocity of light,  M is the mass of the $^{57}$Fe nucleus, and $IS(0)$ is is the temperature-independent part, i.e. the chemical shift. A fit with Eq. (1) to the data of S1 shown in Fig. 6 (a) yields $\Theta$$_D$ = 517(18) K and 545(16) K for 4d and 8h sites, respectively.

\begin{figure}[htp]
\includegraphics[width=0.5\textwidth]{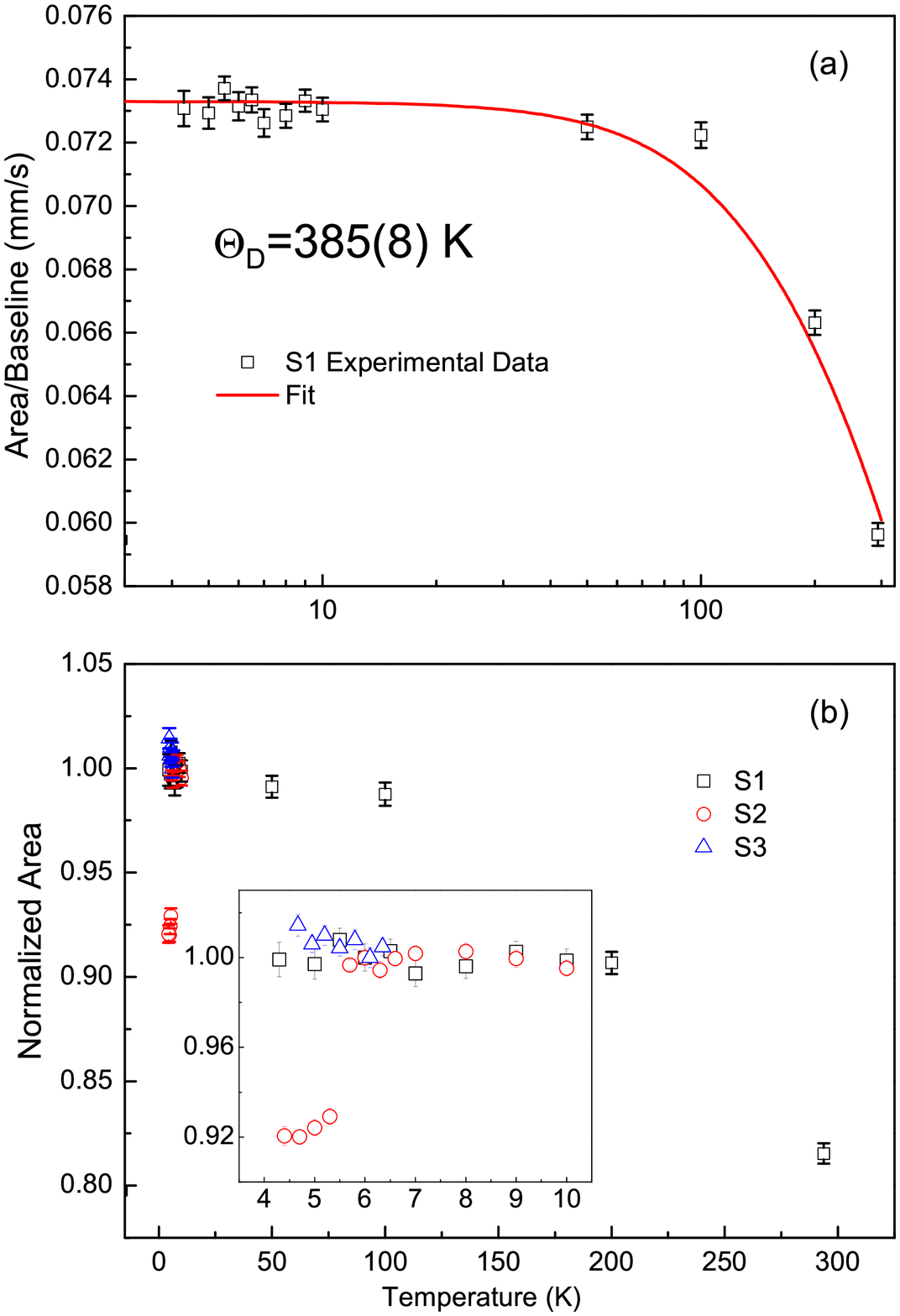}
\caption{\label{fig:epsart}(Color online) (a) Temperature dependence of the spectral area for S1 set of measurements on a semi-log scale. The solid line is a fit to Eq.(2), as explained in the text. (b) Normalized to the values at $\approx 6$ K, temperature dependent spectral area data for all three sets of measurements. Inset: enlarged low temperature part.}
\end{figure}

The quadrupole splittings in Lu$_2$Fe$_3$Si$_5$ at the 4d and 8h sites are 0.50(1) mm/s and 0.19(1) mm/s at 294 K, respectively. The magnitude of the QS is proportional to the $z$ component V$_{zz}$ of the EFG tensor, which is composed of two contributions: (V$_{zz}$)$_{lig}$, from the ligand charges around the M\"ossbauer nucleus, and (V$_{zz}$)$_{val}$, from the valence electrons of M\"ossbauer nucleus. Usually, (V$_{zz}$)$_{lig}$ is small and weakly dependent on temperature, whereas (V$_{zz}$)$_{val}$ is strongly temperature dependent. As can be seen from Fig. 5 (c). the QS of two sites are both almost temperature independent, which indicates the QS is mainly determined by the  contribution from the ligand charge distribution around the $^{57}$Fe sites.

In order to get a better understanding of the electronic origin of EFG at the Fe sites in Lu$_2$Fe$_3$Si$_5$, a first-principles calculation was performed using the full-potential linearized augmented plane wave method as embodied in the WIEN2K \cite{ADD4_CPC1990,ADD5_WIEN2K}. The generalized gradient approximation (GGA) suggested by Perdew, Burke, and Ernzerhof (PBE GGA) \cite{ADD6_PRL1996} was employed for the exchange-correlation effects. Once the electron densities are calculated self-consistently and with high accuracy, the EFG tensor can be obtained from an integral over the non-spherical charge density. The principal component V$_{zz}$ for the 4d site is 2.25$\times$10$^{21}$V/m$^2$ and asymmetry parameter $\eta = (V_{xx} - V_{yy} )/V_{zz} $  = 0.457; the V$_{zz}$ for the 8h site is -1.16$\times$10$^{21}$V/m$^2$ and $\eta$ = 0.307, which qualitatively agree with experimental results. It is found that the $p$-$p$ and $d$-$d$ interactions mainly contribute to the EFG of Lu$_2$Fe$_3$Si$_5$ and $p$- electrons is play a significant role for states far from the Fermi energy whereas the $d$-$d$ interaction dominates around the Fermi energy, which is similar to what is found for the iron-pnictides superconductors \cite{ADD7_PRB2011,ADD8_HI2011}.

The relative areas of the subspectra are determined by the proportion of the Fe atoms on different lattice sites. As mentioned above, for Lu$_2$Fe$_3$Si$_5$, the theoretical relative area of the two subspectra representing iron atoms at the 4d and 8h sites should be 33.3\% and 66.7\%. At RT the relative areas are close to expected value. However, as shown in Fig. 6 (e) as temperature decreases, the relative area values deviate from theoretical values gradually, with the relative area for the 4d site becoming even larger than that for 8h site at low temperature. This phenomenon may be related to the variation of the linewidth, $\Gamma$, for the doublets corresponding to the two sites with reducing temperature. As shown in Fig. 6 (g), the $\Gamma$ of 4d site increases slightly at lower temperature, but the $\Gamma$ of 8h site is almost constant during the whole temperature range. The slightly increase of $\Gamma$ of 4d site suggests the existence of instability at this lattice site at low temperature.

\subsubsection{The spectral area}
In M\"ossbauer studies of superconductivity, the variation of the total spectral area has also been the focus of discussion. There are number of reports showing a decrease of the spectral area near T$_c$ due to the softening of lattice with the opening of superconducting gap in cuprates \cite{1PRL1991,2PRB1992,3PRB1988} but to the best of our knowledge only one report on iron-pnictide superconductors \cite{24SSC2011}. There are two kinds of behavior reported: a rapid decrease near T$_c$ \cite{1PRL1991,4JPCM1993,5JPCM1994}; and a pit-like decrease either around T$_c$ or at higher temperature, serving as a precursor to T$_c$ \cite{2PRB1992}. At the same time,  some publications noted a poor reproducibility of those observations in the cuprates \cite{6SSC1987,25PRB1992}. In our measurements, as can be seen in the Fig. 7, for the S1, we did not observe any abnormal variation around 6.1 K. To make sure we didn't miss any minor variation around T$_c$, we remeasured the M\"ossbauer spectra of the same sample between 4.4 K to 10 K with higher density of the data points around 6.1 K for 48 h. Surprisingly, for data set S2, there is a sharp, 6\%, decrease around 5.5 K. Nevertheless, when we repeated the same measurement again, data set S3, this phenomenon disappeared. The sharp spectral area change seen in data set S2 is, in our opinion, most likely an artifact. The feature in data set S2 occurs resolvably below T$_c$ = 6.1 K with the very sharp and rather large jump occuring between spectra taken at 5.3 K and 5.7 K (with spectra at 5.7 K, 6.0 K, and 6.3 K all having normalized areas near 1.0). If this feature were associated with T$_c$, it should have occured either between 5.7 K and 6.0 K or between 6.0 K and 6.3 K. In addition, there are no corresponding anomalies in data set S2's isomer shift, quadrupole splitting or relative areas (as shown in Fig. 6). A simple explanation for such behavior could be, among others, some mechanical shift/rearrangement of the powder composing the absorber.  This observation gives  a warning that even in the measurement of iron-containing stoichiometric material,  irreproducibilities/artifacts might exist and should be addressed appropriately.

Finally, we also fitted the temperature dependence of area under the two doublets line of S1 measurements with Debye model:
\begin{equation}
  f=exp[\frac{-3E_{\gamma}^2}{k_B\Theta_DMc^2}\{\frac{1}{4}+(\frac{T}{\Theta_D})^2\int_0^{\Theta_D/T}\frac{xdx}{e^x-1}\}],
\end{equation}
where $f$ is the recoilless fraction, which is proportional to the area for thin sample and E$_{\gamma}$ is the $\gamma$-ray energy. This expression also allows to estimate the value of $\Theta_D$. We obtained the $\Theta_D$ = 385(8) K which is very close to the 408 K obtained from the analysis of the low temperature specific heat capacity data and about 140 K less than the value estimated by temperature dependence of IS. A similar difference was found earlier in studies of e.g. FeSe$_{0.5}$Te$_{0.5}$ and $^{57}$Fe- doped YBa$_2$Cu$_3$O$_{6.8}$ compounds \cite{24SSC2011,26SSC1995}. This discrepancy may be explained by the fact the area reflects the average mean-square displacements, while IS related to the mean-square velocity of the M\"ossbauer atom. Both quantities may respond in different ways to the lattice anharmonicities.

\section{Conclusions}
In summary, we performed detailed $^{57}$Fe M\"ossbauer measurements on Lu$_2$Fe$_3$Si$_5$ in the temperature range of 4.4 K to RT. The contributions from two Fe crystallographic sites can be well distinguished by M\"ossbauer spectra. The main contribution of EFG in this compound comes from the lattice anisotropy and the first principles calculations yield the values of EFG that qualitatively agree with the experiment. The Debye temperature was estimated by the temperature dependence of specific heat capacity, spectral area and IS. The $\Theta_D$ obtained from temperature dependence of spectral area and heat capacity are very similar,  but about 140 K smaller than the value estimated by the IS variation with temperature. Additionally, we didn't observe any obvious, abnormal variation of hyperfine parameters around T$_c$. Two possibilities could lead to this result: the opening of the superconducting gap doesn't bring variation of the environment at Fe site at all; the T$_c$ of this system is too low, and M\"ossbauer spectroscopy is not sensitive enough to detect the minute change.

\begin{acknowledgments}
X. M. was supported in part by the China Scholarship Council. The authors (H. P. and F. L.)  gratefully acknowledge the financial support from the National Natural Science Foundation of China under grant No. 11275086. Work at the Ames Laboratory (X. M., S. R., P. C. C. and S. L. B. ) was supported by the US Department of Energy, Basic Energy Sciences, Division of Materials Sciences and Engineering under Contract No. DE-AC02-07CH11358.
\end{acknowledgments}
%
\end{document}